\documentclass[%
 aip,
 amsmath,amssymb,
 10pt,
 twocolumn,
 longbibliography
]{revtex4-1}

\usepackage{graphicx}
\usepackage{dcolumn}
\usepackage{bm}
\usepackage[utf8]{inputenc}
\usepackage[T1]{fontenc}
\usepackage{mathptmx}
\usepackage{color}
\usepackage{chemfig}
\usepackage{textcomp}

\begin{document}

\title{Non-adiabatic quantum dynamics of the H+H$_2^+$ $\rightarrow$ H$_2$+ H$^+$ reaction and isotopic
  variants\\}
\title{Near-resonant effects in the quantum dynamics of   the H+H$_2^+$ $\rightarrow$ H$_2$+ H$^+$ charge transfer reaction and isotopic
  variants\\}


\author{Cristina Sanz-Sanz}
\affiliation{Unidad Asociada UAM-CSIC,
                       Departamento de Qu{\'\i}mica F{\'\i}sica Aplicada, Facultad de
                      Ciencias M-14, Universidad Aut\'onoma de Madrid, 28049, Madrid, Spain}
\author{Alfredo Aguado}
\affiliation{Unidad Asociada UAM-CSIC,
                       Departamento de Qu{\'\i}mica F{\'\i}sica Aplicada, Facultad de
                      Ciencias M-14, Universidad Aut\'onoma de Madrid, 28049, Madrid, Spain}

\author{Octavio Roncero}
\email{octavio.roncero@csic.es}
\affiliation{Instituto de F{\'\i}sica Fundamental (IFF-CSIC),  
                          C.S.I.C., 
                       Serrano 123, 28006 Madrid, Spain}


\begin{abstract}
  The non-adiabatic quantum dynamics of the  H+H$_2^+$ $\rightarrow$ H$_2$+ H$^+$ charge transfer reactions, and some isotopic
  variants, is studied with an accurate wave packet method. A  recently developed $3\times$3 diabatic potential model is used, which
  is based on very accurate {\it ab initio} calculations and
  includes the long-range interactions for ground and excited states. It is found that for initial H$_2^+$(v=0), the quasi-degenerate
  H$_2$(v'=4) non-reactive charge transfer product is enhanced, producing an increase of the reaction probability and cross section.
  It becomes the dominant channel from collision energies above 0.2 eV, producing a ratio, between v'=4 and the rest
  of v's, that increases up to 1 eV.   H+H$_2^+$ $\rightarrow$ H$_2^+$+ H
  exchange reaction channel is nearly negligible, while the reactive and non-reactive charge transfer
  reaction channels are of the same order, except that corresponding to H$_2$(v'=4), and the two charge transfer
  processes compete below 0.2 eV. This enhancement
  is expected to play an important vibrational and isotopic effect that need to be evaluated.
  For the three proton case,
  { the
    problem of the permutation symmetry is discussed when using reactant Jacobi coordinates.
  }
  \vspace*{0.25cm}
\begin{center}
  Accepted in J. Chem. Phys. (2021), \date{\today}
\end{center}
\end{abstract}

\maketitle

\section{Introduction}
Hydrogen is the most abundant element in Universe, with nearly 73\%  of the barionic mass,
and plays a fundamental role in the chemistry of the interstellar medium (ISM). It is at the origin
of the chemical cycles of most of the molecules, which in turn play a fundamental role in the collapse
of molecular clouds to form stars and planetary systems. The most abundant molecular species are
H$_2$ and H$_3^+$, while H$_2^+$, formed by ionization of H$_2$ by cosmic rays, is rapidly destroyed
by the exothermic reaction
\begin{eqnarray}\label{formation-H3+}
  {\rm H}_2 + {\rm H}_2^+ \rightarrow {\rm H}_3^+ + {\rm H}.
\end{eqnarray}
H$_3^+$ is considered as the universal protonator\cite{Oka:12,Oka:13} by producing hydrides when colliding
with atoms and molecules
\cite{Watson:73,Herbst-Klemperer:73,Millar-etal:89,Pagani-etal:92,Tennyson:95,McCall-Oka:00}.
Its  collisions with H$_2$ produce ortho/para transitions of the two
species, and/or to its deuteration when colliding with HD isotopic variant.

In local galaxies,
the formation of H$_2$ is attributed to reactions on cosmic grains and ices\cite{Glover:03,Wakelam-etal:17}, because
the gas phase routes have too low rate constants to reproduce the observed abundances. In Early
Universe, where grains and ices do not exist, one of the key problems is therefore to determine
the processes, and their related rate constants, giving rise to H$_2$ \cite{}. One of this is
the charge transfer reaction
\begin{eqnarray}\label{formation-H2}
  {\rm H} + {\rm H}_2^+ \rightarrow {\rm H}_2 + {\rm H}^+,
\end{eqnarray}
where the reactant and product channels of Eq.~\ref{formation-H2} correspond to the first excited and the ground electronic
states of the H$_3^+$ system, respectively. 

H$_3^+$ plays a fundamental role in astrochemisty and it has been the subject of many studies summarized
in reviews\cite{McNab:95,Tennyson:95,Herbst:00,Oka:12,Gerlich-etal:12,Oka:13} and special issues\cite{H3+special-issue:12,H3+special-issue:19}.
Its infrared spectrum was first  detected in  the laboratory by Oka\cite{Oka:80} and later
in the space\cite{Geballe-Oka:89,Geballe-etal:99,McCall-etal:99,Oka:13}. Since then, H$_3^+$ has become commonly used to probe
spacial conditions, such as a thermometer and a clock of cold molecular clouds\cite{Oka:06}. Its infrared
spectrum has been theoretically characterized with spectroscopic accuracy
\cite{Polyanski2012:5014,Bachorz2009:024105,Velilla2008:084307,Tennyson2017:232001,Furtenbacher2013:5471},
based on highly accurate
potential energy surfaces (PESs), local\cite{Jaquet1998:2837,Tennyson1994:314,Tennyson1995:133} and
global\cite{Cencek1998:2831,Pavanello2012:184303,Mizus2018:1663,Bachorz2009:024105,Velilla2010:387,Rohse1994:2231,Viegas2007:074309,Ghosh2017:074105}.

The H$^+$+ H$_2$ exchange reaction  has
been the subject of many experimental \cite{Schlier-etal:87,Gerlich:90,Gerlich:92,Gerlich-Schlemmer:02,Savin-etal:04,Kusabe-etal:04,Dai-etal:05,Carmona-Novillo-etal:08,Song-etal:05,Urbain-etal:13}
and theoretical \cite{Markovic-Billing:95,Ichihara-etal:96,Last-etal:97,Chajia-Levine:98,%
  Takayanagi-etal:00,Ushakov-etal:01,Ichihara-etal:00,Errea-etal:01,Kamisaka-etal:02,%
  Krstic-Janev:03,Kusabe-etal:04,Chu-Han:05,Gonzalez-Lezana-etal:05,Gonzalez-Lezana-etal:06,Gonzalez-Lezana-Honvault:14,%
  Gonzalez-Lezana-Honvault:17,Sahoo-etal:14,Ghosh-etal:15} studies.
This reaction governs the ortho/para transitions of H$_2$ and is also responsible of the H$_2$ deuteration.
For energies below $\approx$ 1.82 eV, this reaction takes place in the ground singlet state of H$_3^+$.
In the ground  H$_3^+$ electronic state there is a deep insertion well, of $\approx$ 4.6 eV, and the dynamics
is mediated by a dense manifold of resonances, and exact calculations show that the reaction proceeds through
a statistical mechanism\cite{Gonzalez-Lezana-etal:05,Gonzalez-Lezana-etal:06,Gonzalez-Lezana-Honvault:14}.
  Above 1.82 eV collision energy, the H$_2^+$ + H channel becomes accessible,
  opening the charge transfer process, inverse to
the reaction ~\ref{formation-H2}, and has been studied in a broad energy
range\cite{Markovic-Billing:95,Ichihara-etal:96,Last-etal:97,Chajia-Levine:98,%
  Takayanagi-etal:00,Ushakov-etal:01,Ichihara-etal:00,Errea-etal:01,Kamisaka-etal:02,%
  Krstic-Janev:03,Savin-etal:04,Kusabe-etal:04,Chu-Han:05,Urbain-etal:13,Sahoo-etal:14,Ghosh-etal:15}.
Furthermore, the non-adiabatic
transitions in H$_3^+$ were studied recently in photodissociation experiments
by Urbain {\it et al.}\cite{Urbain-etal:19}.

In spite of all these studies on H$_3^+$, reaction~\ref{formation-H2}
has only been studied experimentally by Karpas {\it et al}\cite{Karpas-etal:79},
by McCartney {\it et al.}\cite{McCartney-etal:99}, at energies of 30-100 keV, and
more recently
Andrianarijaona {\it et al}\cite{Andrianarijaona-etal:09,Andrianarijaona-etal:19} studied the H + D$_2^+$ isotopic
variant  in a broader energy range of 0.1-100 eV.
The experimental study of reaction ~\ref{formation-H2} have the difficulty of involving  two radical reactants,
and accurate theoretical simulations are then of high interest to provide realistic rate constants
for the model of Early Universe\cite{Glover:03,Savin-etal:04,Coppola-etal:11,Indriolo-McCall:12,Coppola-etal:13}.
Some approximated theoretical treatments where applied to reaction
~\ref{formation-H2}\cite{Last-etal:97,Kamisaka-etal:02,Krstic:02,Krstic-Janev:03,Errea-etal:05}.
{ Also, an accurate quantum treatment of this reaction was done  very recently
  \cite{Ghosh-etal:21}, during the publication of this work.
}
 In this work we present 
 accurate quantum wave packet calculations  of the H+H$_2^+$ charge transfer (CT) reaction,
 {
   and some isotopic
 variants,
}
at collision energies below 1 eV using
a very accurate set of three coupled PESs recently proposed\cite{Aguado-etal:21}, which include
long range interactions for the ground and excited electronic states. 
The manuscript is distributed as follows. First, in section 2, a brief description of the PESs and
dynamical methods are shown, presenting a description of the permutation symmetry problem. In section 3
the theoretical simulations  are shown and discussed, and, finally, section 4 is devoted to
extract some conclusions, outlining future work.


\section{Methodology}

\subsection{Diabatic electronic representation}
\begin{figure}[hbtp]
  \centering
  \includegraphics[width=0.95\linewidth]{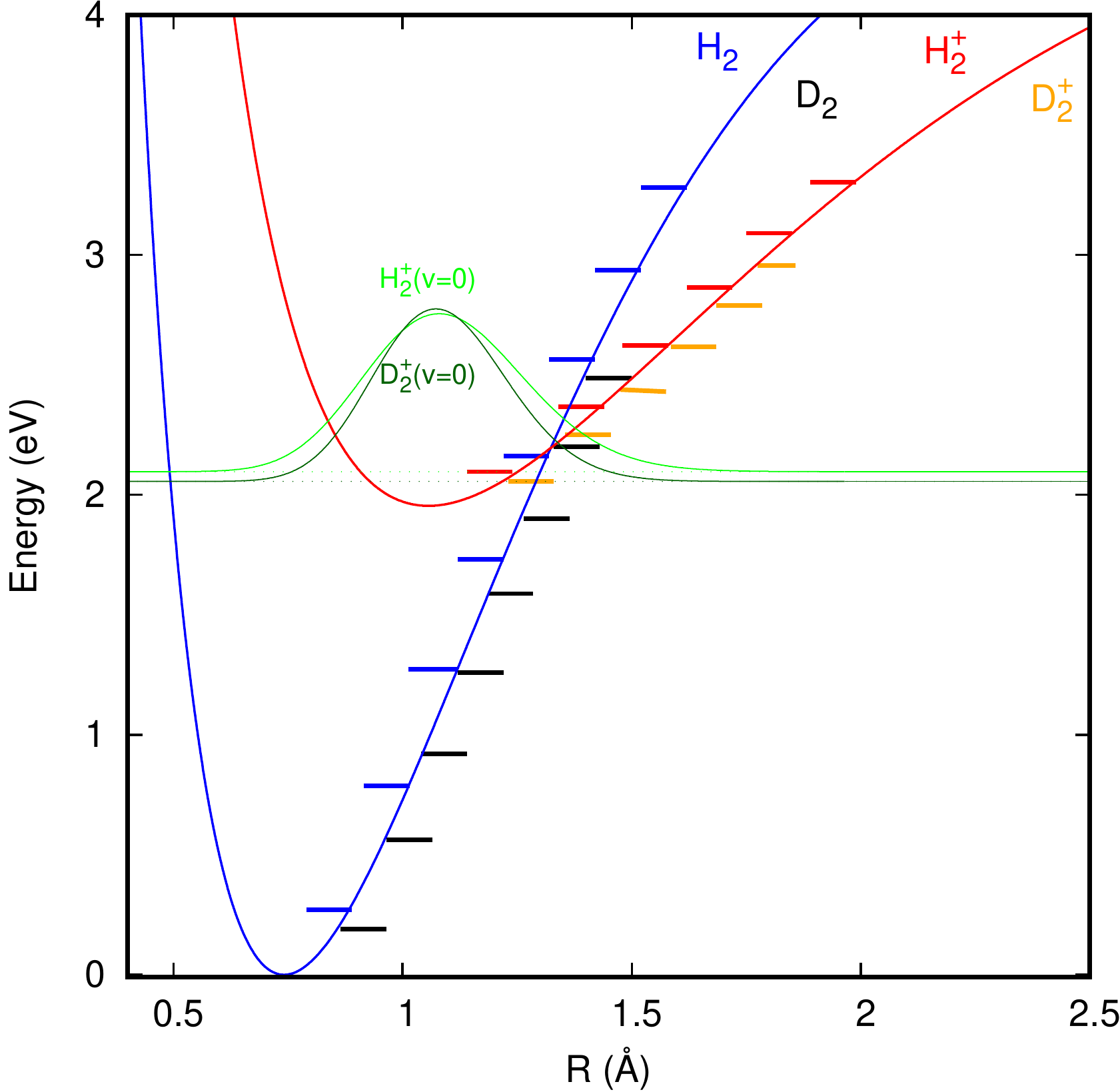}
  \caption{\label{fig:vibrational-levels}
    Potential energy curves and vibrational levels of H$_2$ (D$_2$) and H$_2^+$ (D$_2^+$).
    The  radial part of the nuclear wave functions for H$_2^+$ (D$_2^+$) have been represented in green.
    }
\end{figure}

The potential used is described by a 3$\times$3 diabatic matrix,  each diagonal element
corresponding to a positive charge in each of the nuclei\cite{Aguado-etal:21}. It consists
of a zero-order diatomics-in-molecules  (DIM) matrix\cite{Tully:80},
describing accurately the H$_2$ and H$_2^+$ diatomic fragments
(shown in Fig.~\ref{fig:vibrational-levels}),
plus three-body terms added in the diagonal, and non-diagonal
matrix, $V^{3B}$, as it was proposed by Varandas and co-workers\cite{Viegas2007:074309}.
In this potential the long range term has been improved\cite{Aguado-etal:21} and
included in all the electronic states. The dominant long range terms for H+H$_2^+$ are the charge-induce dipole
and charge-induced quadrupole dispersion interactions, varying  as $R^{-4}$ and $R^{-6}$, respectively,
For H$^+$ + H$_2$ the dominant long range terms are the charge-quadrupole and the charge-induced dipole dispersion energies,
 varying  as $R^{-3}$ and $R^{-4}$, respectively.
The $V^{3B}$ was fitted
to very accurate {\it ab initio} calculations using multi-reference configuration interaction (MRCI) calculations,
performed with the MOLPRO suite of programs\cite{MOLPRO-wires}. A complete basis set (CBS) approximation
was done based on the  aV5Z and aV6Z basis sets of Dunning\cite{Dunning:89}.
These $V^{3B}$ matrix elements change gradually what allows a very accurate
fitting, with a small root mean square error\cite{Aguado-etal:21},
what is an advantage as compared to the fitting of the sharply varying non-adiabatic
matrix elements (NACMEs) required when using an adiabatic representation\cite{Ghosh2017:074105}.

Diabatic representations are generally considered to be approximate, due to the
impossibility to eliminate all NACMEs as a function of all internal coordinates
because of the curl condition\cite{Baer:75,Mead-Truhlar:82}.
The quasi-diabatic representation of the PES used here\cite{Aguado-etal:21} is considered
as a regularization representation\cite{Thiel-Koppel:99,Gomez-Carrasco-etal:06},
in which the singularities present at conical intersections (CI)
are removed exactly\cite{Gomez-Carrasco-etal:06}.
In the vicinity of CIs the calculation of NACMEs presents numerical problems
what adds some error in  the adiabatic representation since the Born-Oppenheimer
is ill-conditioned at CIs\cite{Koppel-cap4-conicalbook:04}.
The resulting NACMEs obtained by Aguado {\it et al.}\cite{Aguado-etal:21}
showed to be in excellent agreement with the {\it ab initio} points.
Moreover, the residual NACMEs
appearing in all diabatic representation are very small in the present case, because the
three-fold basis is rather complete in this system. These residual NACMEs are very small, and only
appreciable far from CIs, where the energy difference among the different electronic states
is so large
that they can be neglected without loss of accuracy.

\begin{figure}[hbtp]
  \centering
 \includegraphics[width=0.95\linewidth]{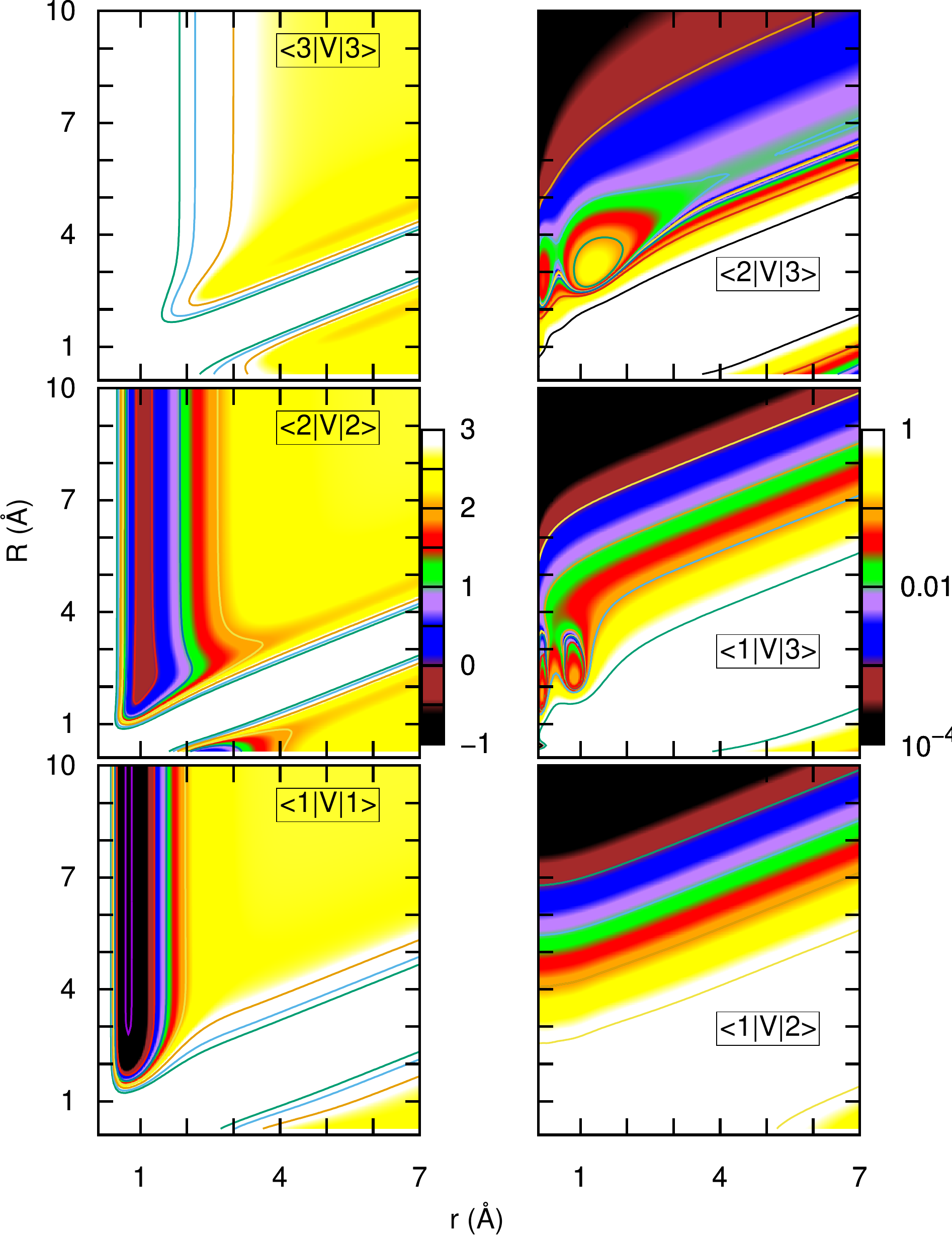}
  \caption{Contour plots of the diagonal (left panels) and non-diagonal (rigth panels) elements of the 3$\times$3
    potential matrix at a Jacobi angle of $0^o$, as a function of the Jacobi distances $r$ and $R$, in the diabatic  representation
    of functions $\phi^1_i$. In the diagonal terms,  the zero of energy is set at the eigen value of H$_2^+(v=0,j=0)$ ground rovibrational
    state, which is at 2.09 eV in Fig.~\ref{fig:vibrational-levels}. The absolute value of nondiagonal terms are plotted in logarithmic scale.
   }
  \label{fig:PES}
\end{figure}

Quasi-diabatic representations are not unique. In Ref. \cite{Aguado-etal:21} it was considered a minimum
basis set formed by 3 functions, denoted $\phi_i$, in which there is a positive charge of nucleus $i$, and
one 1s electron in the remaining nuclei. In this basis set, the excited H$_2^+(^2\Sigma_g^+)$ is a linear combination
of at least two $\phi_i$ functions. A unitary transformation is done, in order to define properly
the H$_2(^1\Sigma_g^+)$ and H$_2^+(^2\Sigma_g^+,^2\Sigma_u^+)$ for a particular (reactant) rearrangement channel, as
\begin{eqnarray}\label{unitary-diabatic-transformation}
  \phi^{1}_1 &=& \phi_1 \nonumber \\
  \phi^{1}_2 &=& {1\over \sqrt{2}} \left( \phi_2 -\phi_3 \right) \\
  \phi^{1}_3 &=& {1\over \sqrt{2}} \left( \phi_2 +\phi_3 \right). \nonumber
\end{eqnarray}
Transforming the 3$\times$3 potential accordingly, the H$_2$ and H$_2^+$ ground
electronic states become diagonal in channel 1, corresponding to the hydrogen 1
 at very long distances from the other two, and the corresponding potentials
are shown in Fig.~\ref{fig:vibrational-levels}. However, on the other two rearrangements,
the potential matrix is non diagonal, and the diabatic couplings are non-zero as can be
seen in the right panels of Fig.~\ref{fig:PES}, where they are represented in Jacobi coordinates, $r$ and $R$
for a collinear configuration. As a consequence, the rovibrational
states of the products are expanded in several electronic states, for H$_2$ and H$_2^+$ as
\begin{eqnarray}
  \phi_{v'j'}^{i=2,3} (r') &=& \sum_k \phi^1_k \, C^{v'j'}_{k}(r') \nonumber\\
 && \label{products-rovibrational-functions}\\
  \phi_{v'j'}^{i=2,3 +} (r') &=& \sum_k \phi^1_k \, C^{v'j'+}_{k}(r'), \nonumber
\end{eqnarray}
respectively, where $r'$ denotes the internuclear distance in the corresponding product Jacobi
coordinates $r'=r_{13}$ or $r'=r_{12}$ for $i$ = 2 and 3, respectively.

The diagonal and non-diagonal elements of the potential matrix in this diabatic representation
are shown in  Fig.~\ref{fig:PES} for a collinear configuration.
For long $R$ and short $r$ values, the diagonal terms represent H$_2(^1\Sigma_g^+)$ (bottom left panel),
the H$_2^+(^2\Sigma_g^+)$ (middle left panel) and  H$_2^+(^2\Sigma_u^+)$ (top left panel), the last being purely
repulsive. The diagonal terms at long $r$ and $R$ values are repulsive, and do not
represent the products channels because the non-diagonal terms (shown in the right panels) are non-zero
and pretty large. In the reactant channels (long $R$ and short $r$ values) the $\langle 1\vert V \vert 2\rangle$ non-diagonal coupling term,
coupling  H$_2^+(^2\Sigma_g^+)$ with  H$_2(^1\Sigma_g^+)$, 
decrease slowly with increasing $R$, being of the order of $\approx$ 1 meV at $R$=7 \AA. This clearly shows
that  this charge transfer coupling extends towards very long $R$ distances.

\subsection{Quantum wave packet calculations}

The reaction dynamics of the H/D collisions with H$_2^+$(v=0,j=0) and H  with D$_2^+$(v=0,j=0) are studied here
 using a quantum wave packet method as implemented
in the MADWAVE3 program\cite{Aguado-etal:97,Paniagua-etal:98,Aguado-etal:03,Gonzalez-Lezana-etal:05,Zanchet-etal:09b}.
For each total angular momentum, $J$,
the state-to-state S-matrix elements are calculated using
a reactant  Jacobi coordinate based method\cite{Gomez-Carrasco-Roncero:06}, generalized
to consider product rovibrational states expanded in several electronic states,
as in Eq.~\ref{products-rovibrational-functions}. Three different processes ($\alpha$ channels)
are distinguished
\begin{eqnarray}\label{processes}
   \begin{array}{r}
     {\rm A} + {\rm B}_2^+(v=0,j=0,J) \longrightarrow  {\rm A}^+ + {\rm B}_2(v',j')\hspace*{1.5cm} \\ \hspace*{1.cm} {\rm non-reactive\, charge\, transfer (NRCT)}\\
       \\
                       \longrightarrow  {\rm AB}^+(v',j') + {\rm B} \hspace*{1.5cm}\\ \hspace*{1.cm} {\rm exchange\, reaction  (ER)}\\
       \\
                       \longrightarrow  {\rm AB}(v',j') + {\rm B}^+\hspace*{1.5cm}\\  \hspace*{1.cm}{\rm reactive\, charge\, transfer (RCT)}\\
   \end{array}
\end{eqnarray}
with $A$= H, D and B$_2$= H$_2$, D$_2$. Channels $\alpha=$ 1 (NRCT) and 3 (RCT) correspond to products in the ground adiabatic
states, either in the reactants or products channels, respectively. Channel $\alpha=2$ (ER) corresponds
to exchange products in the first excited adiabatic state.

\begin{table}[h]
 \caption{\label{wvp-parameters}
   Parameters used in the wave packet calculations in reactant Jacobi coordinates:
   $r_{min} \leq r\leq r_{max}$ is the H$_2$ internuclear distance,
   $R_{min} \leq R\leq R_{max}$ is the distance between H$_2$ center-of-mass and the $A$ 
   atom, $0 \leq \gamma \leq \pi$ is the angle between ${\vec r}$ and ${\vec R}$ vectors. 
}
 \begin{center}
 \begin{tabular}{|cc|}
 \hline 
 $r_{min}$, $r_{max}=$  0.1, 20 \AA & $N_r$=360 \\
 $r_{abs}$=  16 \AA & $\alpha_r$=10$^{-3}$,n=6 \\
$R_{min}$, $R_{max}=$   0.01, 20\AA & $N_R$=360  \\
 $R_{abs}$=  16 \AA  & $\alpha_R$=10$^{-3}$,n= 6\\
$N_\gamma$ = 100 & in $[0,\pi/2]$  \\
$R_0$  = 14 \AA & $E_0,\Delta E$= 0.5,0.5 eV\\
$R_\infty$ = 16 \AA & $R'_\infty$ = 10\AA    \\
 \hline
 \end{tabular}
 \end{center}
 \end{table}
The parameters used in the wave-packet calculations are listed in table~\ref{wvp-parameters}. Briefly,
the initial wave packet corresponds to the product of the H$_2^+$(v=0,j=0) ro-vibrational
eigenfunction, a normalized
Wigner function for $J$ and a real Gaussian function, describing the translation on the $R$
Jacobi distance (between the incomming $A$ atom and the center of mass of H$_2^+$), initially
placed at $R_0$, with a central energy $E_0$ and an energy width $\Delta E$. The flux on individual
rovibrational states are analysed  at $R_\infty$ and $R'_\infty$ for  reactants and products, respectively.
The wave packet is absorbed at the edges of the radial grid by multiplying it by an absorbing function
$e^{\alpha_\rho (\rho-\rho_{abs})^n}$ for $\rho\geq \rho_{abs}$ ($\rho = r$ or $R$) at each Chebyshev iteration,
using a modified Chebyshev propagator\cite{Mandelshtam-Taylor:95}. Using
this propagator only a real wave packet is propagated\cite{Kroes-Neuhauser:96,Chen-Guo:96,Gray-Balint-Kurti:98,Gonzalez-Lezana-etal:05},
and the corresponding kinetic term is evaluated using a sine Fourier transform which ensures the proper
regular behaviour at $R=0$\cite{Lepetit-Lemoine:02}, appearing in this reaction.
Finally, a Gauss-Legendre quadrature is used to describe the Jacobi angle $\cos\gamma= {\bf r}\cdot {\bf R}/rR$,
and the corresponding kinetic terms are evaluated using a Discrete Variable Representation (DVR) method\cite{Roncero-etal:97}.
For $J$=0, about 8$\times$ 10$^4$ iterations are needed to converge the reactions probability down to 0.01 eV. This
long propagation is needed because of the presence of resonant structures that will be commented below. This
is also the situation of higher $J$ until the rotational barrier push the reactive and NRCT probabilities towards
higher energy, what happens at $J$=14 and 17 for H/D + H$_2^+$ cases, respectively. For higher $J$'s, the number of
iterations needed gradually decreases.

 The calculations have been performed for all $J$ up to $J$=15 and $J$=20 for $A$= H and D, respectively.
 For higher values of $J$, wave packet calculations are performed  every 5 $J$ values  up to $J$=60 and $J$=80,
 for $A$= H/D, respectively. A maximum of helicity components $\Omega_{max}$ = 23 is considered
 in the $J>0$ calculations. For intermediate $J$ values, where no wave packet calculations were done,
 an interpolation method based\cite{Zanchet-etal:13} on the $J$-shifting\cite{Bowman:85} approximation
 is performed to evaluate the individual state-to-state $S^2$ matrix elements
 \cite{Zanchet-etal:13}.

 The state-to-state integral cross sections are then evaluated
 using the usual partial wave expansion as
 \begin{eqnarray}
   \sigma_{\alpha v j\rightarrow \alpha' v' j'}(E)&=&
{\pi  k_{\alpha vj}^{-2}\over {(2j+1)}}\sum_{J\Omega\Omega'} (2J+1)
\left\vert S^{J_t}_{\alpha vJ\Omega\rightarrow \alpha' v'J'\Omega'}(E)\right\vert^2    \nonumber\\
&&
\end{eqnarray}

\subsection{Symmetry considerations}

Dealing with two (D+H$_2^+$) or three (H+H$_2^+$) hydrogen atoms, fermions with nuclear spin $i$=1/2,
some considerations about the permutation symmetry should be done. The total wave function
has to be antisymmetric under the exchange of any hydrogen pair.
{
The total
wave function can be factorized as a product of electronic, nuclear spin and rovibrational
components as $\Psi= \Psi_e \Psi_I \Psi_{Rot}$.
The symmetry of the electronic function, $\Psi_e$, is analyzed in the body fixed frame
of the nuclei, and the permutation of any pair of nuclei corresponds to a reflection through
a plane perpendicular to the molecular plane\cite{Sanz-etal:01}. In this case, all singlet
states of H$_3^+$ are totally symmetric.
The nuclear spin functions, $\Psi_I$, are characterized by the total
nuclear spin, $I_2$ and $I_3$, for 2 and 3 hydrogen systems, respectively.
$I_2$= 0 and 1,
and the corresponding functions $\Psi_{I_2}$ are antisymmetric and symmetric
with respect to the permutation of the two hydrogen atoms, respectively.
To make the total wave function antisymmetric,
the corresponding rovibrational
function have to be symmetric and antisymmetric for $I_2$= 0 and 1, respectively. In diatomic
molecules, this corresponds to the usual separation between even and odd rotational
levels, since the symmetry of spherical harmonic is $(-1)^j$,
denoted as para and ortho, respectively.
}

For three hydrogen atoms system, $I_3$= 1/2 and 3/2, and the spin functions can be
written as
$
\vert I_3 M_3; i_2\rangle= \sum_{\sigma m_2} ( i_2, 1/2 \vert m_2, \sigma,I_3 M_3) \vert 1/2,\sigma\rangle \vert i_2,m_2\rangle, 
$
where $\vert i_2,m_2\rangle$ is the spin function of two hydrogen, each one described by $ \vert 1/2,\sigma\rangle$,
and $( \vert )$ are Clebsh-Gordan coefficients. The existing functions are then $\vert 3/2 M_3; 1\rangle$,
$\vert 1/2 M_3; 1\rangle$ and $\vert 1/2 M_3; 0\rangle$, with $I_3$= 3/2, 1/2 and 1/2, respectively.
Considering the $D_{3h}$ permutation-inversion group, the spin functions
with $I_3$=3/2 (ortho) belong to the $A_1$ representation, while with $I_3=$ 1/2 (para) belong
to the $E$ representation.
To build antisymmetric total wave functions, the corresponding rovibrational components
have to belong to $A_2$ (for $I_3$=3/2, ortho) and $E$ (for $I_3$=1/2, para) representations.
It should be noted here,
that the product for $I_3$=1/2, the total symmetry is $E\times E= A_1+A_2+E$,
and only  $A_2$ is anti-symmetric while the other solutions are not physically allowed.

All this said, and neglecting the weak hyperfine nuclear spin rotation coupling, the spin of each hydrogen
atom is conserved, and therefore the transformation
from ortho to para, and viceversa, can only take place by hydrogen exchange, $i.e.$, by a reaction.

When using reactant Jacobi coordinates, the permutation symmetry of the D+H$_2^+$ and H + D$_2^+$ is fully acount for,
while this is not the case for  H+H$_2^+$. In the three identical hydrogen atoms case,
the $A_2$ and $E$ representation has to be considered, separately, while the $A_1$ does not exist. Thus, the
diatomic rotational channels (for reactants and products) included in each representation of the nuclear wave function
are\cite{Miller:69,Honvault-etal:11}

\begin{tabular}{ccccc}
  $E$ &:& even and odd j,j'\, &$\rightarrow$&\, $I_3$=1/2, $I_2$=1,0 (para) of $A_1$ \\
  $A_1$&:& only even j,j' \,  &$\rightarrow$&\, does not exist for $i$=1/2\\
  $A_2$&:& only odd j,j' \,   &$\rightarrow$&\ $I_3$=3/2 (ortho) of $A_1$.\\
\end{tabular}

Clearly the case of three identical hydrogens, with reactants in $j$=0, contains the $E$ and $A_1$
representation of the $D_{3h}$ inversion-permutation group. For the final reactants or products
in even $j'$ also contains $E$ and $A_1$ representations and are therefore difficult to distinguish
when using Jacobi coordinates. Finally, the initial case  $j$=0 and final odd $j'$ (para-to-ortho
transition of dihydrogen molecules/cations) clearly belong to the $E$ representation of H$_3^+$.

Therefore, when using reactant Jacobi coordinates, total integral reaction cross sections can not be
obtained, while the para-ortho state-to-state integral cross section can be obtained. However,
when referring to H+H$_2^+$(v=0,j=0) reaction dynamics we shall include all the cases, physical ($E$)
and non-physical ($A_1$),
to compare with the D+H$_2^+$(v=0,j=0).

\section{Results}

\begin{figure*}[!bth]
  \centering
  \includegraphics[width=0.75\linewidth]{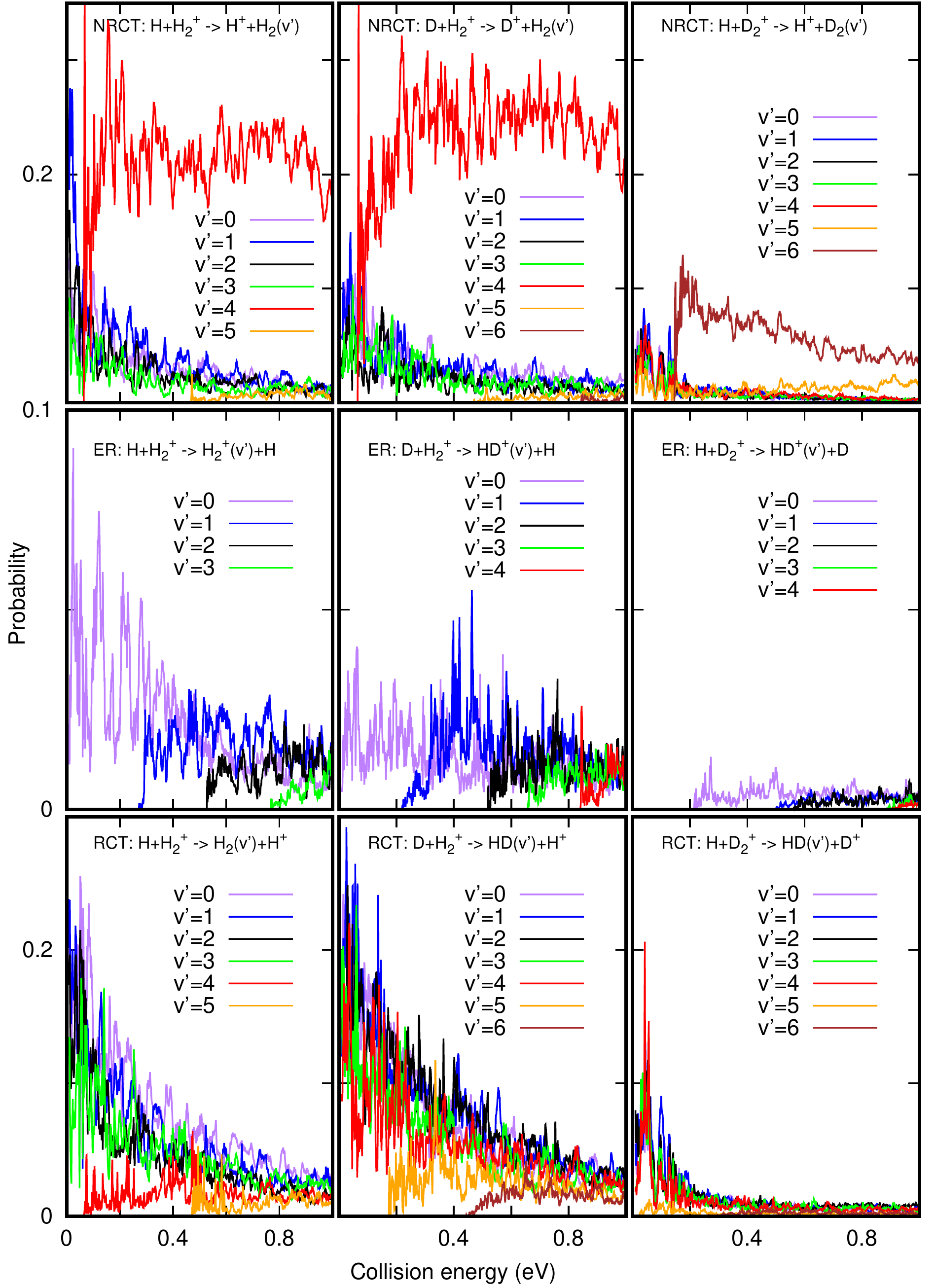}
  \caption{Vibrationally resolved reaction probability for H+H$_2^+$(v=0,j=0) (left panels), D+H$_2^+$(v=0,j=0) (middle panels),
    and H +D$_2^+$(v=0) (right panels), 
    towards the inelastic charge transfer, NRCT (H$^+$+H$_2$, D$^+$+H$_2$ and H$^+$+D$_2$, in the top panels),
    the exchange reactive channel, ER (H+H$_2^+$, H+HD$^+$ and D +HD$^+$, in the middle panels) and reactive charge transfer, RCT, channels
   (H$^+$+H$_2$, H$^+$ +HD and D$^+$ +HD, in the bottom panels).}
  \label{fig:probJ0}
\end{figure*}

The reaction probabilities obtained for the three reactions under study for $J$=0 are
shown in Fig.~\ref{fig:probJ0}, separated for the three different processes of Eq.~\ref{processes}
and summing over all rotational states of products. The reactions for H+H$_2^+$ (left panels), D+H$_2^+$ (middle panels)
and H+D$_2^+$ (right panels) show  similar patterns, and each mechanism is discussed separately below.

 For the three reactions, the NRCT process
 (in the top panels of Fig.~\ref{fig:probJ0}) is open from zero collision energy.
 As displayed in Fig.~\ref{fig:vibrational-levels}, the H$_2^+$(v=0) classical turning
 points appear at shorter distances than  the crossing point between  the H$_2^+$ and H$_2$ potential curves.
 However,  the vibrational ground state of H$_2^+$(v=0) and D$_2^+$(v=0) have  significant
 amplitude at the region of the crossing, see  Fig.~\ref {fig:vibrational-levels}.
 This situation allows the electronic charge transfer in the same channel 
 for long $R$ distances, because the electronic coupling becomes effective (see Fig.~\ref{fig:PES}).

 This 
 is particularly evident for H$_2$(v'=4) which is nearly resonant with H$_2^+$(v=0)
 (just 0.064eV above), presenting a large vibrational overlap, what yields to
 a strong electronic transition to H$_2$(v'=4). The ratio between the final
 probability in H$_2$(v'=4) and all other channels increases with collision energy. This clearly
 explains the experimental results\cite{Karpas-etal:79},
 who found that charge transfer is the dominant mechanism, rather than hydrogen exchange or
 complex formation with scrambling. These two mechanisms, associated to ER and
 RCT, can compete at the lower energies considered here, as discussed below.

 For D$_2^+$(v=0), however, the energy differences with  D$_2$(v'=5,6) are larger, of $\approx$
 0.15 eV, and the amplitude of the  D$_2^+$(v=0) at the electronic crossing is 
 lower (see Fig.~\ref{fig:vibrational-levels}). All these make less effective the electronic coupling.
 As a consequence, all the final vibrational channels have lower probability.
 {
   In spite 
   of the larger energy spacing, the  D$_2$(v'=6) level shows to be considerably more
   populated that any other vibrational level of D$_2$, in clear analogy with the situation
   of H$_2$.
 }

 \begin{figure*}[bhtp]
  \centering
  \includegraphics[width=0.75\linewidth]{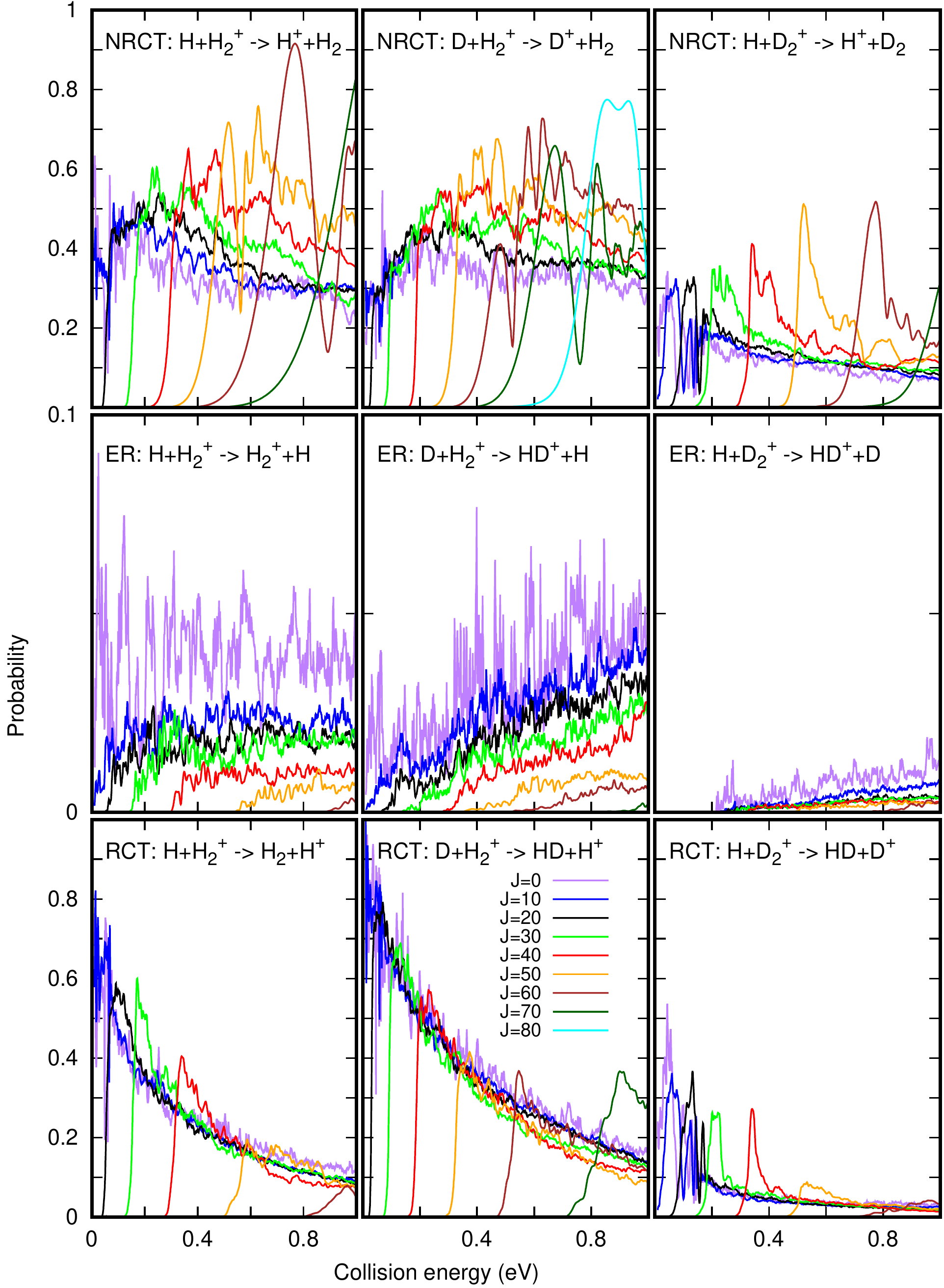}
  \caption{Total reaction probability at different total angular momentum, $J$,
    for H+H$_2^+$(v=0,j=0) (left panels), D+H$_2^+$(v=0,j=0) (middle panels) and H+D$_2^+$(v=0,j=0),
    towards the inelastic charge transfer, NRCT (H$^+$+H$_2$, D$^+$+H$_2$ and H$^+$ + D$_2$ products in the top panels),
    the reactive channel, ER (H+H$_2^+$, H+HD$^+$ and D+HD$^+$ in the middle panels) and reactive charge transfer, RCT, channels
   (H$^+$+H$_2$, H$^+$ +HD and D$^+$+HD in the bottom panels).}
  \label{fig:probJ}
\end{figure*}

 The exchange reaction mechanism, ER in the middle panels of  Fig.~\ref{fig:probJ0},
 are also non zero, even when it is less probable than the other two charge transfer
 mechanisms. The adiabatic potential for the first excited electronic state 
 presents
 a rather high reaction barrier for the exchange. Therefore,  ER proceeds in a two
 step mechanism. First, there is a  transition to the ground
 electronic state, with a deep well, explaining the resonance structure present
 in the ER probabilities. This is followed by a second transition back to the
 excited H + H$_2^+$/H+DH$^+$/D+DH$^+$ channel, in each case respectively,
 giving rise to non-zero ER probabilities, lower
 than the RCT probabilities because there is a lower density of final states for H$_2^+$/HD$^+$ channels.
 The probability for the ER channel presents a clear decreases when the mass of reactants increase,
  from H+H$_2^+$, to D+H$_2^+$ and H +D$_2^+$.

 The RCT mechanism (see bottom panels of Fig.~\ref{fig:probJ0})
 consists of two steps, a transition to the ground electronic state
 and the reactive exchange of a hydrogen atom. Except for H$_2$(v'=4) (or D$_2$(v'=6)),
 the RCT probabilities
 are very similar to those of the NRCT channel. This could be explained
 by the near statistical mechanism of the reaction in the ground adiabatic
 state\cite{Gonzalez-Lezana-etal:05,Gonzalez-Lezana-etal:06,Carmona-Novillo-etal:08,Honvault-etal:11},
 that is, the reaction is mediated by resonances originated by
 the deep insertion well of the ground electronic state of H$_3^+$. At collision energies
 below 1 eV these resonances are long lived
 and the reaction cross section and the final distribution of products
 are well described by statistical methods\cite{Gonzalez-Lezana-etal:06,Honvault-etal:11}.
 At the higher energies considered
 here (about 1.8 eV above the ground H$^+$+H$_2$ threshold), the resonance 
 persist and a pseudo-statistical behaviour may be expected.
 This explains the resonance structure of all reaction probabilities
 in Fig.~\ref{fig:probJ0}.

 { 
 Three main differences are found with the previous results recently
 reported by Ghosh {\it et al.}\cite{Ghosh-etal:21}.
 First, the results of Ref.\cite{Ghosh-etal:21} do not show the dense manifold
 of narrow resonances shown here. Also
 their probabilities do not show a progressive increase with decreasing collision energy,
 as the present ones do, associated to the long range interactions.
 Finally, the results of Ghosh {\it et al.}\cite{Ghosh-etal:21} do show a much lower
 increase of the NRCT for H$_2$(v'=4), and instead their probabilities for
 H$_2$(v'=3) are of the same order of v'=4.
 All these differences are attributed to the use of different PESs, and,
 in particular, to the electronic couplings
 producing the charge transfer. It is worth noting here,
 the coupling at relatively long distance plays an important
 role in the nearly resonant enhancement of the  CT for H$_2$(v'=4) obtained here.
 On the contrary, Ghosh {\it et al.}\cite{Ghosh-etal:21}
 used a relatively short initial distance of $\approx$ 5.6 \AA\ as compared to $R_0$=14 \AA\ used here.
}

The total reaction probabilities for each of the mechanisms and different values of total angular
momentum, $J$, are shown in Fig.~\ref{fig:probJ} for H+H$_2^+$,D+H$_2^+$
{ and H+D$_2^+$ }
reactions.
All the probabilities shift towards higher energies
with increasing $J$. However, there are important differences. First, D+H$_2^+$ shows smaller shifts,
simply because the effective rotational barrier decreases,  because the reduced mass is larger,
$\mu\approx$ 1, 4/5 and 2/3 amu for D+H$_2^+$, H+D$_2^+$ and H+H$_2^+$, respectively.
The NRCT channel also shows a smaller
shift with $J$, what is attributed to the long-range character of the electronic couplings, as described above.
Moreover, this also explains why the NRCT probabilities increase near the rotational threshold while
for ER and RCT the probabilities clearly decrease with $J$ at the threshold.
As the rotational barrier increases, it closes the access to reaction, while the NCR process is still
possible due to long range non-adiabatic couplings, which are effective because the quassi-degeneracy
of the H$_2^+$(v=0) and H$_2$(v'=4) levels,
{ 
  or equivalently D$_2^+$(v=0) and D$_2$(v'=6).
}
This behavior also produces
an enhancement of the NRCT mechanism as compared to ER and RCT channels,
except for low collisions, where RCT mechanism
dominates.

{
 The opacity function, {\it i.e.} the reaction probabilities versus $J$, reported by Ghosh {\it et al.}\cite{Ghosh-etal:21}
 shows a sudden drop between 0.3 and 0.4 eV at $J$=12. However, in the present results
 the rotational threshold at 0.3 eV
 is approximately at $J$=40.  A continuous and gradual increase of the rotational
 threshold  is found here at all collision energies of Fig.~\ref{fig:probJ}. This difference with previous
 results\cite{Ghosh-etal:21} is 
 attributed to the long-range behaviour of the electronic couplings and the longer distances
 considered in the dynamical calculations, as discussed above.
}

\begin{figure}[hbtp]
  \centering
  \includegraphics[width=0.95\linewidth]{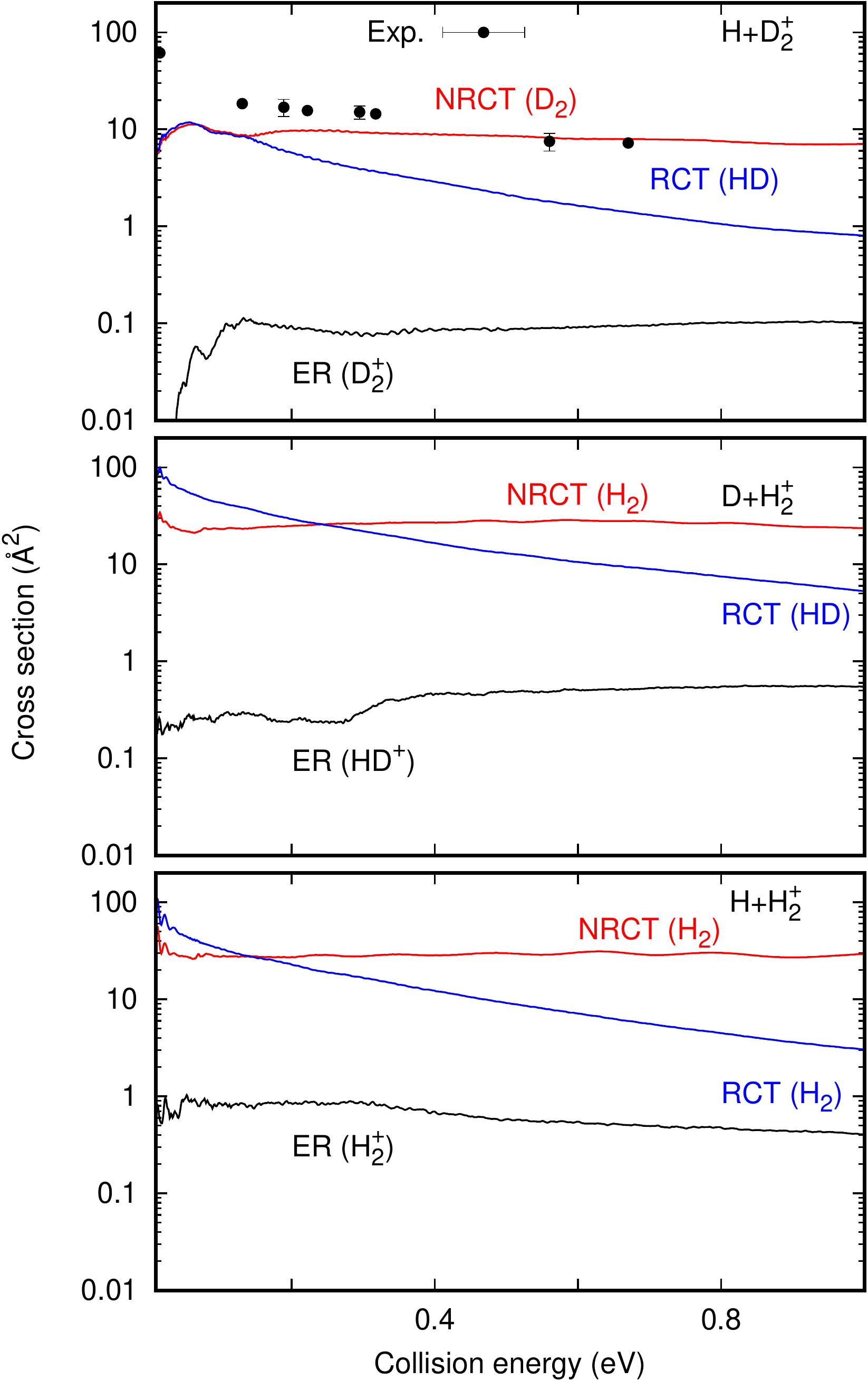}
  \caption{Total integral reactive cross sections for  H+H$_2^+$(v=0,j=0) (bottom panel),
     D+H$_2^+$(v=0,j=0) (middle panel) and H+D$_2^+$(v=0,j=0) (top panel),
    towards the inelastic charge transfer, NRCT (H$^+$+H$_2$, D$^+$+H$_2$  and H$^+$+D$_2$ products),
    the reactive channel, ER (H+H$_2^+$, H+HD$^+$ and D+HD$^+$ products)
    and reactive charge transfer, RCT, channels
    (H$^+$+H$_2$, H$^+$ +HD and D$^+$ +HD). In the top panel the points correpond to the experimental
    values from Ref.~\cite{Andrianarijaona-etal:09,Andrianarijaona-etal:19}
    \label{fig:total-sigma}
    }
\end{figure}
The total integral cross section for each of the processes are shown in Fig.~\ref{fig:total-sigma}
for the  H+H$_2^+$(v=0,j=0) (bottom panel), D+H$_2^+$(v=0,j=0) (middle panel)
{
  and H+D$_2^+$(v=0,j=0) (top panel)
}
collisions.
At  1 eV, $\sigma^{NRCT} > \sigma^{RCT} > \sigma^{ER}$, but the ratio changes with collision
  energy. The exchange reaction, ER, is always lower, since it involves 2 electronic transitions
  and the H$_2^+$/HD$^+$ products have a lower density as compared to H$_2$/HD.
  The NRCT/ER ratio is nearly constant with collision energy. On the contrary NRCT/RCT varies
  considerably, being nearly
  { 
  1 for collision energies near  $0.2$ eV,
  }
  and increases to
{ 
  10/4/9 at 1.eV, for H+H$_2^+$/ D+H$_2^+$/ H+D$_2^+$ respectively.
}
  This ratio seems to increase
  with energy,
  {
    in agreement with the experiments performed by Karpas {\it et al.}\cite{Karpas-etal:79},
  who found that 
  the CT dominates the H+H$_2^+$ reaction.
  }

  At low collision energies,
  {
    before the dominant NRCT resonant H$_2$(v'=4) or D$_2$(v'=6) channel opens,
    RCT and NRCT cross sections are of the same order, and in most cases
    the RCT process dominates. In this region
    the dynamics is dominated by  a statistical process:
  }
  After a first
  electronic transition, the system gets trapped in the long-lived resonances
  originated by the deep H$_3^+$ well of the ground electronic state. At
  these long-lived resonances, energy transfer among all internal degrees of freedom becomes
  very effective
  and the products are formed proportionally with the
  density of states. Thus for H$_2^+$, with three identical H$^+$ + H$_2$ charge transfer
  products, the NRCT/RCT reaches a factor between 0.5 and 1. For D$^+$ + H$_2$, this
  factor decreases, since HD$^+$ RCT products have larger density of states than H$^+_2$+ D.
  {
    Finally, for H+D$_2^+$ the ratio of the density of products states is reversed
    becoming denser for the inelastic D$_2$ channel than for the reactive HD one. This
    explains why here the RCT is nearly identical to NRCT in this case.
  }

  { The NRCT for the  H + D$_2^+$(v=0,j=0)  reaction is compared 
    with the experimental work of Andrianarijaona {\it et al.}\cite{Andrianarijaona-etal:09,Andrianarijaona-etal:19}
    in the top panel of Fig.~\ref{fig:total-sigma}.
    In these measurements of the H + D$_2^+$ reactions, only H$^+$ products are detected,
    so that their cross section 
    only corresponds to the NRCT process. The good agreement obtained with the present
    results, specially at energies of $\approx$ 0.6 eV, already demonstrates the
    adequacy of the simulations done here. It should be noted, however, that
    the D$_2^+$ reactants are produced with some vibrational excitation. This could
    explain why at lower energies the agreement is worse.
    Some studies on the vibrational effects
    are now under way.
    }

\begin{figure}[hbtp]
  \centering
  \includegraphics[width=0.95\linewidth]{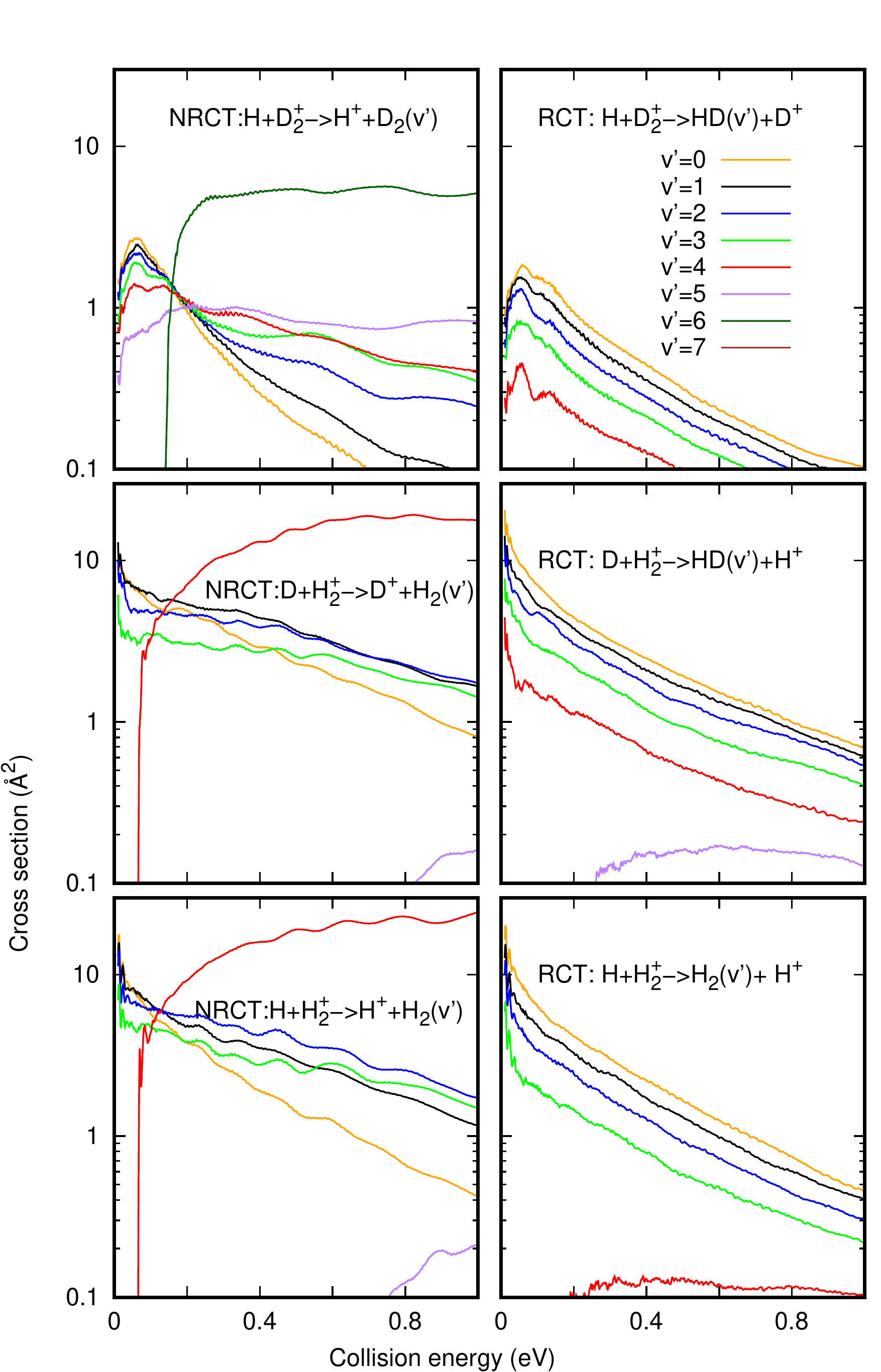}
  \caption{Vibrationally resolved  charge transfer cross sections for
    H+H$_2^+$(v=0,j=0)$\rightarrow$ H$^+$ + H$_2$(v')/ H$_2$(v')+ H$^+$ (bottom panels)
    D+H$_2^+$(v=0,j=0)$\rightarrow$ D$^+$ + H$_2$(v')/HD(v')+H$^+$ (middle panels)
    and H+D$_2^+$(v=0,j=0)$\rightarrow$ H$^+$ + D$_2$(v')/HD(v')+D$^+$ (top panels)
    , for the NRCT (left panels) and RCT (right panels) processes.}
  \label{fig:vib-CT-sigma}
\end{figure}
  The NRCT dominates the high energy region because the nearly resonant conditions
  between H$_2$(v'=4) and the H$_2^+$(v=0) levels
  {
    (or D$_2^+$(v=0) and D$_2$(v'=6)),
  }
  as discussed for the reaction probabilities.
  The vibrationally resolved NRCT cross sections, in Fig.~\ref{fig:vib-CT-sigma},
  also show this effect.
  In the first two  reactions, below  the H$_2$(v'=4) channel opens, at 0.064 eV, H$_2$(v')
  are formed progresively in ascending order, v'=0,1,2 and 3.
  There are small isotopic effects, but in general the NRCT shows a typical decreasing
  behaviour associated to exothermic reactions. At energies above 0.064 eV, H$_2$(v'=4) opens,
  and the cross section for this channels increases becoming the dominant channel
  at energies above 0.2 eV.
  {
    A similar situation holds for H+D$_2^+$(v=0), but replacing  H$_2$(v'=4) by D$_2$(v'=6).
    }
 
  The vibrational resolved cross sections of Fig.~\ref{fig:vib-CT-sigma} are in qualitative
  agreement with the results in Fig. 7 by Krstic\cite{Krstic:02}, which includes both NRCT and
  RCT processes calculated in a broader energy range using a close coupling method
  based on the infinite order sudden approximation (IOSA) with Delves hyperspherical
  coordinates. In Krstic work,   $v'$=4  cross-section is always the most important,
  but it becomes of the same order than that for $v'$=3 at
  collision energies of 0.2 eV. Below 0.2 eV  all $\sigma_{v=0\rightarrow v'}$ CT cross sections
  are of the order of 10 \AA$^2$\ \cite{Krstic:02}, and for all $v'\neq$ 4 decreases monotonically
  until 3 eV. At 1 eV the cross section for $v'=3$ is of the order of 1 \AA$^2$,  similar
  to the present results of NRCT+RCT. The differences arise about the position of the crossing
  between $v'$=4 and the other $v'$, and on the relative magnitude of the cross section associated
  to different v's. Since the method used here is more accurate, it may be concluded
  that at energies below 1 eV the present results are more accurate.

  The calculations of Last {\it et al.}\cite{Last-etal:97} were done from 0.06 to 0.21 eV
  using a  quantum method based on negative imaginary potential combined
  with a variational quantum method in a L$^2$
  basis set. Their NRCT final vibrational distributions pick at $v'$=3, instead of 4. However,
  they used the helicity decoupling approximation, {\it i.e.} they did not considered the coupling
  between different helicities $\Omega$. In this case, this approximation is not appropriate, because
  there are strong couplings between the different resonances with different $\Omega$ value
  originated by the deep well in the ground electronic state. However, the reason why their
  results are picked at $v'$ =3 and not $v'=4$ should be related to the use of different
  potentials for the H$_2$ and H$_2^+$ fragments or to the extension of non-adiabatic couplings.

\begin{figure}[hbtp]
  \centering
  \includegraphics[width=0.95\linewidth]{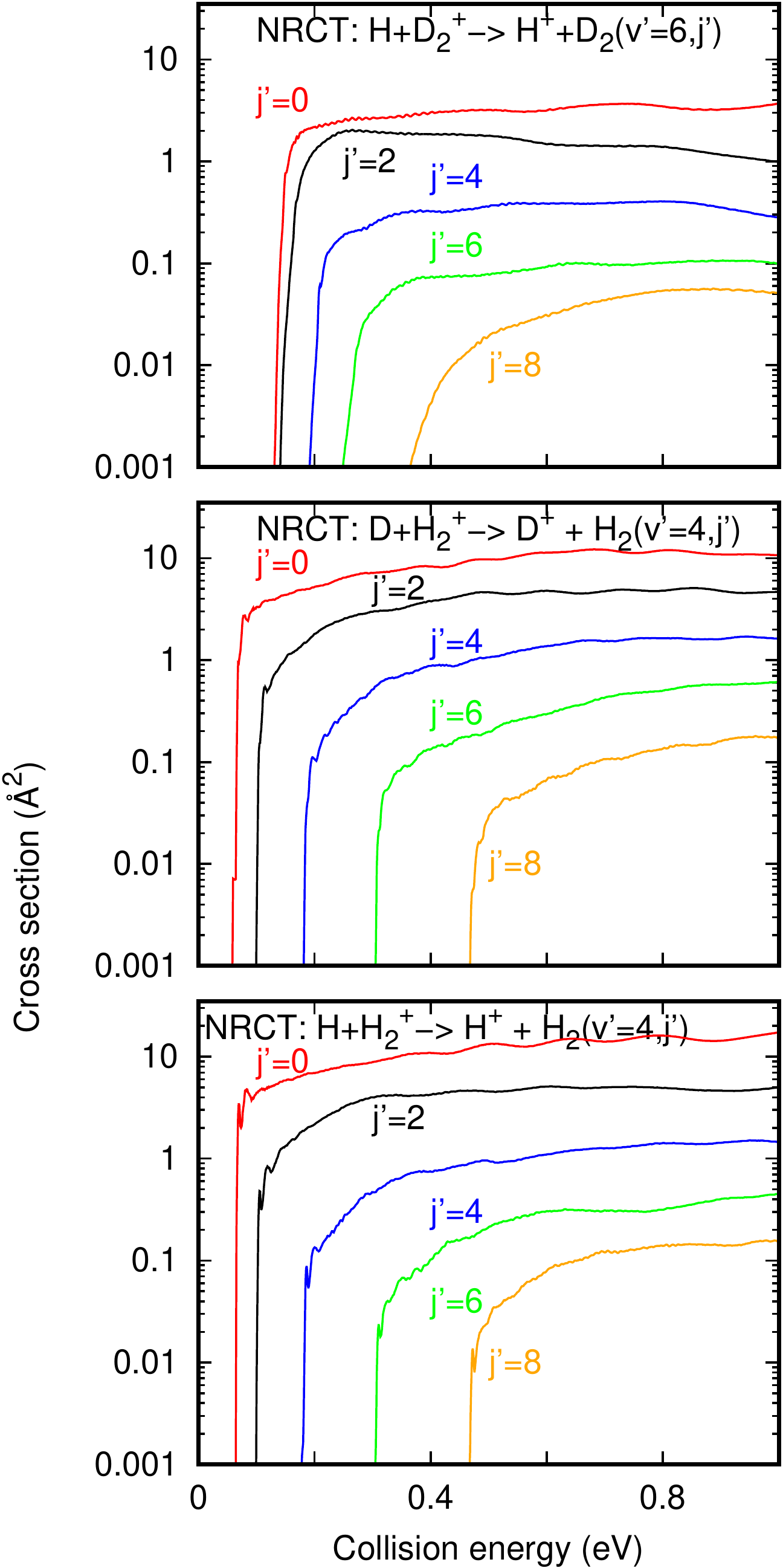}
  \caption{Rotationally resolved  NRCT cross sections for
    H+H$_2^+$(v=0,j=0)$\rightarrow$ H$^+$ + H$_2$(v'=4,j') (bottom panel),
     D+H$_2^+$(v=0,j=0)$\rightarrow$ D$^+$ + H$_2$(v'=4,j') (middle panel)
    and H+D$_2^+$(v=0,j=0)$\rightarrow$ H$^+$ + D$_2$(v'=6,j') (top panel).}
  \label{fig:rot-NRCT-sigma}
\end{figure}

The rotationally resolved NRCT cross sections for  H/D+H$_2^+$(v=0,j=0)$\rightarrow$ H$^+$/D$^+$ + H$_2$(v'=4,j')
and H+D$_2^+$(v=0,j=0) $\rightarrow$ H$^+$ + D$_2$(v'=6,j')
collisions, in Fig.~\ref{fig:rot-NRCT-sigma}, clearly show that j'=0 is the dominant final rotational
state, which is the closer to the H$_2^+$(v=0,j=0) or D$_2^+$(v=0,j=0) initial state. j'=2/j'=0 ratio is about a factor
of 1/4 for collision energies above 0.3 eV, and this ratio decreases with rotational excitation of
CT products. Last  {\it et al.}\cite{Last-etal:97} also reported the maximum of the state-to-state
cross section for final  H$_2$(v'=4,j'=0) state, but less than a factor of 2 as compared to v'=3,j'=4.

Finally, it should be noted that while the results
are accurate for D+H$_2^+$ and H+D$_2^+$, for H+H$_2^+$ we are including also the non-physical $A_1$ irreducible
representation. While some work is now in progress to include the full permutation symmetry for  H+H$_2^+$,
we can already use as exact the para-to-ortho state-to-state cross sections.
Moreover,  because of the high propensity of the NRCT to final H$_2$(v'=0,j'=0), some conclusions can be extracted.
Thus, the present results for H+H$_2^+$ asses that the NRCT channel is dominant for energies above 0.2 eV up to 1eV,
in clear agreement with the results of Karpas {\it et al.}\cite{Karpas-etal:79}.

\section{Conclusions}

In this work accurate quantum calculations have been presented for
the H/D+H$_2^+$
{ and H + D$_2^+$
}
non-adiabatic charge transfer reactions using
an accurate 3$\times$3 diabatic potential matrix recently
proposed\cite{Aguado-etal:21}, which includes 
long range interactions. It is found that the dominant
channel corresponds to the resonant non-reactive charge transfer
from H$_2^+$(v=0) to  H$_2$(v'=4) and D$_2^+$(v=0) to  D$_2$(v'=6), which is enhanced by their
nearly resonant energies and the long-range dependence of
the non-adiabatic couplings.
{
  This enhancement is based on rather general features of the
  H$_3^+$ asymptotic features, namely the electronic crossing
  and the energy difference of the vibrational states 
  of the neutral and cation systems, and are not expected
  to depend on the accuracy of the PESs used in this work. The only
  requirement is to include the electronic coupling existing at rather long
  distances between the reactants, which is described very accurately
  by the PESs used in this work.
  }

For D+H$_2^+$ { and H+D$_2^+$}
state-to-state cross sections are presented
for the first time. For H+H$_2^+$  some caution must be paid because
the permutation symmetry is not fully accounted, while some work
is now-a-days in progress to include the permutation among the three
fermions. In spite of that, the present results show a good qualitative
agreement with the experimental data of Karpas {\it et al.}\cite{Karpas-etal:79},
who also reported that the CT channels dominat at $\approx$ 1.8 eV.
{
  The rather good agreement with the NRCT cross sections measured
  for H+D$_2^+$ by Andrianarijaona {\it et al.}\cite{Andrianarijaona-etal:09,Andrianarijaona-etal:19}
  already demonstrates the adequacy of the simulations presented here for E$>$0.5 eV.
  However, a further study on the vibrational effect of the D$_2^+$ reactants
  should be done to understand the behavior at lower energies.
}
It is also necessary to extend the present
calculations to other rovibrational states and isotopic variants to provide reaction rate constants
of interest in astrophysical models of the Early Universe.

\section{Acknowledgements}
{
  We acknowledge Dr. Andrianarijaona for providing us with the experimental
  data and very interesting discussions of their results.
}
The research leading to these results has received fundings from
MICIU (Spain) under grant FIS2017-83473-C2.
We also acknowledge computing time at Finisterre (CESGA) and Marenostrum (BSC)
under RES computational grants ACCT-2019-3-0004 and  AECT-2020-1-0003, and CCC (UAM).

\section{Data availability}
The data that support the findings of this study are available from the corresponding author
upon reasonable request.
 
%

\end{document}